# Knowledge workers' collaborative learning behavior modeling in an organizational social network

Przemysław Różewski[1] Jarosław Jankowski[1] Piotr Bródka[2] Radosław Michalski[2]

Faculty of Computer Science and Information Technology[1]
West Pomeranian University of Technology
prozewski@wi.zut.edu.pl, jjankowski@wi.zut.edu.pl

Institute of Informatics, Wrocław University of Technology[2]
Wrocław, Poland
radoslaw.michalski@pwr.edu.pl, piotr.brodka@pwr. edu.pl, kazienko@pwr.edu.pl

Abstract:

Computations related to learning processes within an organizational social network area require some network model preparation and specific algorithms in order to implement human behaviors in simulated environments. The proposals in this research model of collaborative learning in an organizational social network are based on knowledge resource distribution through the establishment of a knowledge flow. The nodes, which represent knowledge workers, contain information about workers' social and cognitive abilities. Moreover, the workers are described by their set of competences, their skill level, and the collaborative learning behavior that can be detected through knowledge flow analysis. The proposed approach assumes that an increase in workers' competence is a result of collaborative learning. In other words, collaborative learning can be analyzed as a process of knowledge flow that is being broadcast in a network. In order to create a more effective organizational social network for co-learning, the authors found the best strategies for knowledge facilitator, knowledge collector, and expert roles allocation. Special attention is paid to the process of knowledge flow in the community of practice. Acceleration within the community of practice happens when knowledge flows more effectively between community members. The presented procedure makes it possible to add new ties to the community of practice in order to influence community members' competences. Both the proposed allocation and acceleration approaches were confirmed through simulations.

Keywords: organizational social network, computational models, collaborative learning behavior, knowledge flow, community of practice

.

# 1 Introduction

There is no doubt that the concept of collaboration is closely related to learning. The collaboration process, in which people interact, employs self-critiquing (reflection); inquiry and arguing skills are a solid base for the (social) constructivism pedagogy that is commonly utilized in modern companies (Schaf et al., 2009). Today, almost every company wants to become a knowledge-creating company. Knowledge management pioneer Nonaka (Nonaka et al., 2000) claims that making personal knowledge available to others through social networks is the central activity of a knowledge-creating company. It takes place continuously and at all levels of an organization.

In the knowledge management area, the main focus rests on information technologies (IT). The problem of how knowledge can be shared effectively among workers using organizational social relationships has been marginalized (Dong et al., 2012). Prior research on knowledge management shows that the proper arrangement of organizational social relationships significantly impacts the efficiency of knowledge sharing. Researchers have noticed a move from a technological-based knowledge management strategy to a socialization-based knowledge management strategy as companies seek to more effectively facilitate knowledge sharing.

Recent works bring some insight to the problem. Long and Qing-hong's (2014) study investigated how to divide users into collaborative learning groups. They utilized the users' educational interests to group them into customized clusters. In each cluster, a genetic algorithm was adopted for collaborative learning group division based on a user's knowledge level in order to approximate the optimal development of a collaborative learning group. Another approach to the problem of efficient design and the use of knowledge flows in order to maximize worker knowledge level (over a planning horizon) through sharing in different organizational environments was presented by Dong et al. (2012). In this approach, organizations that support multiple skills and have workers with varying levels of knowledge in these skills were examined. The algorithm developed identified the best set of knowledge transfers in each period in order to maximize the total weighted knowledge level of a given organization over a planning horizon. As a result, the mixed integer programming model and its related heuristics were formulated to facilitate the systematic analysis and understanding of effective knowledge flows. Neither of these approaches included a community-of-practice component or roles in the knowledge flow.

Depending on the analysis concept, there are different approaches to collaboration network analysis (Różewski, 2010). The queuing theory can be used to efficiently optimize a telecommunications network (Różewski and Ciszczyk, 2009). In such a situation, the node represents various computer stations that are able to signal regeneration or data distribution. Another approach to collaboration network analysis is from the workflow point of view (Wang et al., 2006). In this context the network's unit is a task and we are looking for workload optimization. Moreover, the nodes correspond to workstations with assigned technological operations. The last approach to collaboration network analysis, which is used in this article, treats the collaboration network as an agent's network (knowledge network). The analysis can then use social network analysis (Newman, 2003), network game theory (Jackson, 2008), competence set theory (Yu and Zhang, 1990), or ontology theory (Gomez-Perez et al., 2004). Thus, in the network competence set, knowledge/information object or concept, and knowledge resources all circulate. The control parameters are communication efficiency, completeness, and credibility. The emergency and synergy are the work paradigm. The node represents a social agent.

The insertion point of the research arises with the dilemma of homogeneous and heterogeneous group creation in the network. Bekele and Graf (2006) show that heterogeneity can increase learning effects in collaborative learning. High-ability students help low-ability students, as a result the former can remember the knowledge they have acquired longer. One of the ways to create heterogeneous groups is by taking learning styles into consideration. Felder and Silverman's (1988) model provides four dimensions: perception, reception, processing, and understanding. A similar framework was designed by Conole et al. (2004). Based on a bipolar set of learning styles from the literature, algorithms for heterogeneous group creation are proposed (Bernacki and Kozierkiewicz-Hetmańska, 2014). Moreover, the topic of recommended learning material that is suitable for students' characteristics, needs, and preferences was presented in Kozierkiewicz-Hetmańska's work (2011).

In this article, the research problem addresses collaborative learning through knowledge flows in the design of an organization. Knowledge flows are the most important elements of the collaborative learning process in an organizational social network. For this reason, we want to understand exactly how they move through the network. Besides the cognitive and social abilities of the knowledge workers and their relationships, the knowledge that flows is the main influencer on the workers' collaborative learning process. In addition, an effective collaborative learning process results in competence development. Moreover, we assume that knowledge flow is more intense in a community of practice. As a result, in the presented research, we want to establish different methods to make knowledge flows more efficient with respect to the different roles in the network and the community of practice. In the proposition, a number of concepts are combined into one model and all of them will be described in the upcoming sections of the article.

The approach presented in this article extends the available models toward the concept of knowledge workers, who are described by information concerning their competences (in vector format) and mask data structures, which reflect a worker's ability to labor in a specific area. Moreover, knowledge diffusion in the network is achieved by knowledge resource broadcasting. The workers' collaborative learning behavior is described through a computational model and allows for the analysis of different worker configurations and relationship statuses.

This article is divided into four parts. The following section covers the theoretical background related to the problem. In particular, attention is paid to competence development in an organization, knowledge flow in the description of communities of practice, and the collaborative learning development process. The model for a knowledge network in an organization is described in Section 3. The model is based on the formalization of knowledge resources that are transferred by knowledge flows throughout the network. Section 4 describes the method for role allocation in an organizational social network. The roles involved are those of knowledge facilitator, knowledge collector, and expert. The next section analyzes the problem of community of practice acceleration through the addition of new relationships.

# 2 Theoretical Background

## 2.1 Competences in an Organization

There are a number of ways to understand the concept of competence depending on the origin of the field of science or humanities being referenced. The French word "compétence" was originally used to describe the capability of performing a task in the context of vocational training (Romainville, 1996). Later on, the word found its place in general education, where it was mainly related to the "ability" or "potential" to act effectively in a certain situation. Perrenoud (1997) claimed that competence was not only limited to the knowledge of how to do something but also reflected the ability to apply this knowledge effectively in different situations. Grant and Young (2010) analyzed and summarized the skills and knowledge approach to competence.

The requirements for the development of a competence-based approach come from staff development and deployment; job analysis reveals the need for new approaches to knowledge modeling in organizations (Radevski et al., 2003). In modern companies, the competence-based approach is a main component of employment planning, recruitment, training, increasing work efficiency, personal development, and managing key competences. Draganidis et al.'s (2008) study showed that a competence-based approach can identify the skills, knowledge, behaviors, and capabilities needed to meet current and future personnel selection needs that are in alignment with various strategic and organizational priorities. Moreover, a competence-based approach can focus on the individual as well as group development plans in order to eliminate the gap between the competences needed for a project, job role, or enterprise strategy and those that are currently available. Sanchez (2004) reported some challenging issues that must be addressed with a competence-based approach, including: the development and use of a consistent set of concepts and vocabulary for describing competences, the classification of different types and levels of activities within organizations that collectively contribute to achieving competence, and the articulation of interactions between different types and levels of organizational activities that are critical in the processes of competence building and leveraging.

The representation of competence in the information system is based on the ontology framework (García-Barriocanal et al., 2012; Draganidis et al., 2008; Jussupova-Mariethoz and Probst, 2007). Macris et al. (2008) described why the ontological structure is appropriate for competence processing. The most important consideration is that ontology allows for the definition of an organization-wide role structure based on the competences required by different job functions and organizational positions. Moreover, ontology helps identify the competences required to perform the various activities involved in each business process and assigns roles to process these activities based on the competences. Additionally, ontology is a base for the identification of the competences that have been acquired in the organization and for the assignment of users to roles through competence matching.

In the literature, two different base concepts of competence coexist (Bass et al., 2008). An interesting discussion of this issue can be found in McHenry and Strønen (2008), who concluded that the first concept defines competence by targeting individual workers while the second one defines competence by the results of the work produced. We analyzed this issue based on McHenry and Strønen's work. The first competence concept focused on individual competences and took the workers' attributes as the starting point for discussing competence. The workers' competence value was treated as a stock that could be developed through training and validated in "objective" rating schedules. In the second concept, competence was conceptualized as a characteristic of organizations where human competences are seen as one of the available resources.

## 2.2   Knowledge Flow in Communities of Practice

According to Kirschnera and Laib (2007), a community of practice is a process in which social learning occurs because the people who participate in the process have a common interest in some subject or problem and are willing to collaborate over an extended period with others who have this same interest. From another perceptive, communities of practice are groups of people who share a concern or passion for something they do and who learn how to do it better as they interact regularly (Wenger et al., 2002). The results of communities of practice members' collaboration are ideas, the finding of solutions, and the building of a repository of knowledge that changes each member's competence. Moreover, in many industry sectors the community of practice is recognized as a key to improving performance (Abel, 2008).

In the work of Zhuge et al. (2005) we found a number of definitions related to the previously discussed issue of knowledge flow in communities of practice. Knowledge flow is the process of passing knowledge within a team. In other words, knowledge flow is a process of knowledge interchange in a cooperative team (Guo et al., 2005). A similar definition was created by Li (2007): knowledge flow is the process of knowledge diffusion, knowledge transfer, knowledge sharing, and relevant knowledge increase caused by the aforementioned items, which results from interaction between different actors, including the organization and the individual. A knowledge flow begins and ends at a knowledge node. A knowledge node is either a team member or a role that can generate, process, and deliver knowledge. A knowledge flow network is made up of knowledge flows and knowledge nodes.

In modern companies, knowledge flows networks are used to facilitate knowledge sharing. The research carried out by Cowan and Jonard (2004) presents the impact of different types of network structures in the context of knowledge diffusion across organizations based on a simulation. The knowledge flow network has to satisfy the following predetermined conditions in order to create effective flows (Zhuge et al., 2005): knowledge nodes in the network use similar intelligence to acquire, use, and create knowledge; knowledge nodes share knowledge autonomously; knowledge nodes share knowledge without reserve; and the team is cooperative, small, and flat within the organization. Moreover, the geographical, cognitive, and social distance is an important consideration for knowledge flows between individuals (Østergaard, 2009). Guo et al. (2005) describe why knowledge passing and sharing only happens when trust is present.

The communities of practice supported by effective knowledge flows can provide task-relevant knowledge to community members that helps them fulfill their knowledge needs quickly and effectively (Liu et al., 2013).

## 2.3 Collaborative Learning Development

Collaborative learning is a learning method that helps workers study through intra-group collaboration and competition between groups (Long and Qing-hong, 2014). Due to the largely Internet-based and intercultural workplace of many professionals, the collaborative learning process is migrating toward computer-supported collaborative learning (Popov et al., 2014; Colace et al., 2006). Knowledge workers, the members of the collaborative learning community, may participate in various collaboration activities in different ways based on their competences (Kolodner, 2007). At the organization level, the group composition, group size, collaborative media, and learning tasks may differ (Rummel and Spada, 2005).

The classic learning process in universities is teacher-centered and, due to cost limitations and organizational obstacles, it cannot be directly implemented in companies. However, collaborative learning supports a company's needs for training and worker self-learning. According to Kuljis and Lees (2002), the principles of collaborative learning are based upon a learner-centered model that treats the learner as an active participant. The members of the cooperative group are encouraged to carry on deeper conversations, create multiple perspectives, and develop reliable arguments. This is the main reason why collaborative groups facilitate greater cognitive development than what the same individuals can achieve while working alone (Hutchins, 1995). The higher levels of human–human interaction are a solid foundation for collaboration in an organization (Schaf et al., 2009).

In order to develop collaborative learning in the company network system, we have analyzed individual learning interest and people's knowledge level as users, as along with their quantifications, and have established a user model (Long and Qing-hong, 2014). This approach is similar to community building. In order to make a collaborative learning network effective, all groups need to coordinate their efforts and resources in effective ways (Kwon, 2014). The task of building an effective collaborative learning network is composed of two sub-problems (Long and Qing-hong, 2014): how to choose and quantify the proper features to build a user model for a collaborative learning network and how to divide the users into optimal teams in order to achieve their learning goals. Research shows that workers need unique group regulatory behavior, because sharing common ground is paramount for effective collaboration with other group members (Kwon, 2014). Moreover, the thoughtful design of a collaborative learning network must include scaffolding to encourage the desired approaches and behavior (Willey and Gardner, 2012). Furthermore, any culturally diverse members of the group need to overcome an additional level of complexity due to culture-related differences (Popov et al., 2014). Other issues related to building a collaborative learning network include the cognitive, motivational, and socio-emotional challenges that are experienced in collaborative learning, understanding how conflict emerges, and what students' emotional reactions and interpretations are (Näykki et al., 2014; Ayoko et al., 2008).

From the technological side, collaborative learning activities can be realized through the following modes (Zhao and Zhang, 2009): face-to-face collaborative learning, asynchronous collaborative learning, asynchronous distributed collaborative learning, or synchronous distributed collaborative learning. It should be noted that another research problem is the optimal selection of an information system for different modes of the IT market (Colace et al., 2014).

# 3   Model of a Knowledge Network in an Organization

## 3.1   Knowledge Worker

From the market's point of view, a company's global objective is to maintain a position in the market. In a knowledge-based economy, increasing the company's intellectual capital is a primary element of this strategy (López-Ruiz et al., 2014; Nemetz, 2006). Moreover, from the knowledge perspective, the organization's knowledge worker competences and any related core competences are an important part of intellectual capital (Ulrich, 1998). Core competences are abilities that are unique to the company in the market (Ligen and Zhenlin, 2010). However, due to tough competition, the competitive advantage comes from not only owning these kinds of competences, but also having high levels of them, or at least higher levels than a competitor has. The key to the successful operation of an organization is to effectively manage the process of transferring knowledge, which allows the company to use its assets in the most effective way (Dong et al., 2012; Różewski et al., 2013).

Let us assume that organization $X$ is composed of knowledge workers determined by index $i$, where $I = \{i : i \in N_+\}$. All the knowledge workers in the knowledge-based organization are characterized by a set of competences. Knowledge workers enhance their competences by taking part in projects and cooperating with other workers (who are willing to share their knowledge and who have higher competences), by attending training courses, and through self-study (Różewski et al., 2013). All organizational competences are related to a worker's knowledge set and are stored in a competence bank.

The structure of a competence bank is developed by an organization's management and plays a strategic role in the organization. A competence bank is represented by the vector that contains all of the competences in the organization: $\overline{Cb} = [cb_1, cb_2, ..., cb_n, ..., cb_N]$. Each vector element, $cb_n$, represents the maximal value of competence $n$ among all employees. Some competence values may be equal to zero. In that case, a strategic goal for the organization would be to increase the value of this competence. If we assume that the set $Cb$ consists of all the elements of vector $\overline{Cb}$ and that set $Cc$ represents core competences, then $Cc \subseteq Cb$ is a subset of the organization's competences. The core competences are the most important part of an organization's intellectual capital. More information about core competence can be found in Bonjour and Micaelli (2010).

The level of competence $n$ for worker $i$ is calculated by the audit procedure:

$$c_{n,i} = audit(n,i) \tag{1}$$

The audit procedure is based on various methods and techniques for competence analysis (Grant and Young, 2010). From the point of view of the competence audit, each competence has a name and a set of attributes that define it. Each of the attributes for a given employee is evaluated is some way (e.g., questionnaire, interview) (Koeppen et al., 2008). The aggregated attributes allow us to calculate a worker's competence level.

Every worker, $i$, possesses a competence set characterized by a competence vector $\overline{C}_i = [c_{1,i}, c_{2,i}, ..., c_{n,i}, ..., c_{N,i}]$, where $n$ is the number of organizational competences in the competence bank and a value of $c_{n,i} \geq 0$ represents its initial estimate based on the audit procedure. Moreover, the value of $c_{n,i}$ can be changed through knowledge transfer, learning, forgetting, and other knowledge-related processes.

In the literature (Farrington-Darby and Wilson, 2006), the level of competence is normalized and associated with the expertise of a given employee, e.g., novice (0–0.2), initiate (0.2–0.4), apprentice (0.4–0.6), journeyman (0.6–0.8), expert (0.8–1), and master (1). However, in the discussed model, the level of competence does not have an upper limitation due to the open nature of the knowledge process in an organization. In some cases, the competence level can be transformed to a linguistic variable in order to obtain some kind of Likert's scale (e.g., based on the fuzzy approach [Guillaume et al., 2014]). Additionally, an employee with more competence (expert) within a given domain is skilled, competent, and thinks in qualitatively different ways than novices (Anderson, 2000).

The workers' personal communication abilities in a social network are characterized by their cognitive and social abilities. The cognitive ability for node $v_i$ is $o_i \in <0,1>$. The highest $o_i$ and the fastest actor behind $i$ is able to learn and acquire knowledge from others in order to increase his/her knowledge level. The social ability for node $v_i$ is $s_i \in <0,1>$. The highest $s_i$ and the fastest actor behind $i$ is able to teach others. This means that such an individual has the social skills to adapt (personalize) communication to the recipient (Xu et al., 2005).

In addition to his/her competence set, every worker is defined by the purpose of his/her action. In the proposed model, the current area of interest is defined by the selection vector $\overline{\alpha}_i$. The definition of a selection vector for worker $i$ follows $\overline{\alpha}^i = [\alpha_1^i, \alpha_2^i, ..., \alpha_n^i, ..., \alpha_N^i]$, $\alpha_n^i \in \{0,1\}$. Applying a selection vector on a competence vector yields a selected worker's set of active competences. If $\alpha_n^i = 0$, then competence $n$ is outside the scope for the current time. All communication with coworkers and other activities are filtered by the selection vector.

## 3.2  Network Definition

We can distinguish between different levels of networks in an organization. However, all of these layers should be reduced to a one-dimensional network in order to make processing more effective. For example, every employee is related to his/her peers through social, work-related, and other kinds of relationships. Furthermore, the communication-based social network is created from the data collected within the organization, such as e-mail logs, phone call records, surveys, and other sources (Michalski-Kazienko, 2014). A number of research papers have covered the issue of social networks by mining from different organizational sources and metadata (Kilduff and Tsai, 2003), information diffusion in multilayer networks (Michalski et. al, 2013) or application of branching processes (Jankowski et al., 2013). In our approach, the organizational social network is a network structure that was created from the social, organizational, operational, and other layers of various companies. More information about the different layers of company integration can be found in Michalski and Kazienko (2014) and Maier (2007).

In order to estimate the strength of existing relationships between employees, we have to integrate all the networks in a common structure. In most cases, we need to assess relationship strength through the analysis of different types of relationships between employees. Moreover, due to the complex nature of organizational relationships, the resulting network will be very dense. All layers are based on the same set of nodes, where every node represents a knowledge worker. The graphs with multiple edge types are denoted as multilayer graphs but can be transformed into a single-layer undirected graph (Boden et al., 2012).

The organization network for organization $X$ is an undirected graph without self-loops $G^X = (V, E, f)$, where $V = \{v_i\}$ is a set of nodes representing knowledge worker $i$, $E \subseteq V \times V$ is the set of edges representing a symmetrical relationship between nodes (knowledge workers), and $f : E \rightarrow \Re^+ \cup \{0\}$ is a variable edge-labeling function. The function $f(v_x, v_y)$ returns the weight of relation between nodes $v_x$ and $v_y$.

The neighborhood of a given knowledge worker (node $i_1$) is a set $\Gamma_{i_1} = \left\{ v_{i_2} : v_{i_2} \in V, e(v_{i_1}, v_{i_2}) = 1 \right\}$. The $e(v_{i_1}, v_{i_2})$ is a binary variable for $v_{i_1}, v_{i_2} \in V$ and $i_1, i_2 \in I$. If a connection between $v_{i_1}$ and $v_{i_2}$ exists, then $e(v_{i_1}, v_{i_2}) = 1$; otherwise, $e(v_{i_1}, v_{i_2}) = 0$.

## 3.3 Knowledge Resource Broadcast in a Network

The traditional approach to knowledge resources includes the following elements in this group (based on Zhen et al., 2011): design cases, patents, technical standards, design formulae, design rules, software, and experts. In our approach, we focused on the communication between knowledge workers and did not model the content of the knowledge resources. The value of specific knowledge resources is determined by their impact on the competence set in a given resource's consumer. As a result, in order to increase the value of a specific knowledge worker's competence, he/she has to receive proper knowledge resources. The knowledge resources are transferred or exchanged during the employee's collaborations. In the competence context, knowledge resources can be possessed, transferred, acquired, developed, and stored.

We can estimate the amount of competence available in knowledge resources based on the competences of the person who created the resource and his/her social abilities to teach. The change in competence value is influenced by the recipient's cognitive abilities, the social ability of the sender, the recipient's competences with regard to the knowledge resource, and the knowledge resource itself. We assume that $R_i = \left\{ \bar{r}_1^i, \bar{r}_2^i, ..., \bar{r}_m^i, ..., \bar{r}_M^i \right\}$, $\bar{r}_m^i = [r_{m,1}^i, r_{m,2}^i, ..., r_{m,n}^i, ..., r_{m,N}^i]$, and $r_{m,n}^i \geq 0$ represent all the knowledge resources that the knowledge worker $i$ can create and send over the collaboration network.

The incoming knowledge resources are processed in the Resources Processing Block (Fig. 1) only if the resource creator's level of competency is higher than the resource receiver's level. The processing operation represents the knowledge acquisition process. As a result, the value of competence is changing. Every node can generate resources that are immediately transmitted to all of the connected nodes and placed in the Resource Processing Block. Now, let us introduce the time index: $T = 1, ..., t, t+1, ..., t*$.

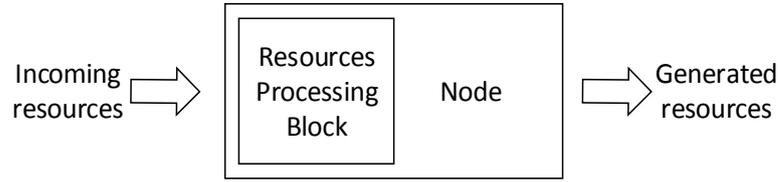

**Fig. 1.** Node Structure

Generally speaking, the generic processing mechanism is as follows:

*Given:*

- Initial level of competence $c_i$ for node $i$

- Selection vector for node $\alpha^i$

- Set of incoming resources $R = \{r^y\}, y \in \Gamma_i$, placed in the Resource Processing Block in node $i$

- $o_i$ cognitive abilities of knowledge worker from node $i$

*To calculate:*

$$c_i[t+1] = c_i[t] + \psi(r^y, \alpha^i) \qquad (2)$$

where function $\psi(r^y, \alpha^i)$ represents the processing of selected elements (based on $\alpha^i$) of the incoming resource $r^y$, according to node $i$ characteristics.

*Constraints:*

$$c_y[t] > c_i[t] \qquad (3)$$

Condition (3) assumes that the competence of sending node $y$ is greater than the receiving of node $i$.

In the proposed model, the employee distributed newly created knowledge resources to all connected employees. The knowledge resources were broadcast according to the following procedure:

1. knowledge resource creation,

2. knowledge resource transmission, and

3. knowledge resource assimilation.

### 3.3.1 Knowledge Resource Creation

Let us assume that knowledge resources are created by employee $A$. Every element of vector $\bar{r}_m^A$ has the following definition:

$$r_{m,n}^A = s_A \cdot c_{n,A} \cdot \alpha_n^A \tag{4}$$

The quality of the developed knowledge resources depends on worker $A$'s social ability to teach $s_A$ and his/her competence vector $\bar{C}_A$. Moreover, the selection vector for knowledge resources $\bar{\alpha}^A$ helps locate the selected employee $A$'s specific set of competences in the knowledge resources. Some part of the knowledge related to competence $c_{n,A}$ is stored in the resources. If $\alpha_n^A = 0$, then competence $n$ is outside of the knowledge resources. The knowledge resources can be saved and stored in the knowledge repository for future use.

### 3.3.2 Knowledge Resource Transmission

The knowledge resource created by worker $A$ in node $v_A$ is transmitted to all of his/her neighbors (from set $\Gamma_A$). The knowledge network reflects all relationships in the organization in the form of weighted edges. As a result, the knowledge resource transmitted from node $v_A$ to node $v_B$ must take the relationship's value into account: $\bar{r}_m^{A \rightarrow B} = \bar{r}_m^A \cdot f(v_A, v_B)$ for every element of vector $\bar{r}_m^A$.

### 3.3.3 Knowledge Resource Assimilation

The knowledge resource (developed by $A$) is processed by employee $B$ in his/her Resource Processing Block. Every component of the knowledge resource is analyzed separately and is processed only if employee $A$'s competency level in this area is higher than employee $B$'s. Every element of new competence vector $\bar{C}_B$ has the following definition:

$$c_{n,B}[t+1] = \begin{cases} (1-\beta^B) \cdot c_{n,B}[t] + \alpha_n^B \cdot r_{m,n}^{A \to B}, & if\ c_{n,A}[t] > c_{n,B}[t] \\ (1-\beta^B)c_{n,B}[t], otherwise \end{cases} \qquad (5)$$

where $\beta^B$ is the forgetting factor of employee $B$ and refers to the ratio of the lost competence level after he/she has forgotten knowledge for some reason (similar to Qu et al., 2010). Employee $B$ processes knowledge resources in order to increase the value of his/her own competences. The final value of competence, $c_{n,B}$, is dependent on employee $B$'s cognitive skills, $o_B$, and resources content, $r_{m,n}^A$, as well as employee $B$'s previous competences vector, $\overline{C}_B'$. The selection vector $\overline{\alpha}^B$ corresponding to employee $B$ is of interest. If $\alpha_n^B = 0$, then competence $n$ is not processed by employee $B$.

# 4 Roles in Knowledge Network Allocation

## 4.1 Roles in the Knowledge Network

From some perspectives, only negative and positive role identification is required in a knowledge network (Brendel and Krawczyk, 2008). In this case, we focused on knowledge development and opposite knowledge deterioration and disintegration. However, in the proposed model, we analyzed the different roles related to knowledge processing.

A broad overview of roles in a knowledge network can be found in Maier's work (2007). Maier distinguished the following roles:

- knowledge manager (builds a knowledge culture, designs a knowledge management strategy, acquires knowledge, measures the value of intangible assets),
- subject matter specialist/expert (quality assurer; knowledge editor; very knowledgeable about certain domain areas, subjects, or processes; tends to have very focused and concentrated experience),
- knowledge administrator (helps others capture, store, and maintain knowledge independent of the domain),
- knowledge base administrator (repository maintaining),
- knowledge broker (helps participants locate the knowledge or experts needed),
- boundary spanner (maintains contacts between experts in different fields),

- knowledge sponsor/skeptic (excited/unenthusiastic about the idea of knowledge management),
- community manager (management of [virtual] community or networks of experts in organizations), and
- mentor/coach (responsible for the development of new talent and competences).

In addition, Awazu (2004) introduced gatekeepers (control the knowledge that enters or leaves a network) and bridges (connecting people who do not share common backgrounds, skills, or experiences). The paper by Boari and Riboldazzi (2014) adds two other roles: representative (communicates information to or negotiates exchanges with outsiders) and liaison (links distinct groups without any prior allegiance to each other).

In our approach, we focused on three roles: knowledge facilitator, knowledge collector, and expert. All of these individuals integrate many of the roles presented earlier. The knowledge facilitator plays the role of the knowledge sponsor, administrator, and broker, who maintains contact between the workers (experts) in different fields and facilitates a faster flow of knowledge in the network. The expert (mentor/coach) introduces new knowledge into the network. As a result, the knowledge flow in the network can be redesigned. The knowledge collector is responsible for knowledge transfer to the company's repositories and plays the role of knowledge administrator and gatekeeper. One important issue we tend to overlook is the problem of management. In our opinion, the management issue will be important after the knowledge flow has been optimized.

## 4.2  Role Allocation

In real-world situations, information about a worker's cognitive and social abilities, as well as his/her level of competence, is difficult and costly to determine. For this reason, in the role allocation process, we focused on the network structure and the social characteristics of the network. Let us define the actions in time with relation to the nodes that accept the new role:

- The node $v_f$, which plays the role of knowledge facilitator, has to increase its relationship power by value $\lambda$: $f[t+1](v_l, \{\Gamma_l\}) = f[t](v_l, \{\Gamma_l\}) \cdot \lambda$ for $l \in I$.
- The knowledge collector role is node $v_c$ with the biggest incoming ratio: $\sum_c \bar{r}^{i \to c}$ for $c, i \in I$.

- The expert is node $v_e$ with an explicitly higher value of competence in the network: $c_{n,e}[t+1] >> c_{n,e}[t]$ for $e \in I, n = 1...N$.

In order to select nodes for specific roles, we created different strategies for node selection (Table 1). All presented strategies $S^T = \{S1, S2, S3, S4, S5, S6\}$ take into consideration a specific set of network characteristics. Strategies S1–S4 rely on well-known metrics from Social Network Analysis (Newman, 2003).

**Table 1.** Summary of information about strategies for ranking development

| No. | Name | Description | Main Concept | Best Value |
|-----|------|-------------|--------------|------------|
| S1 | Random | All nodes are selected based on randomness. | Randomness | – |
| S2 | Degree | The nodes are ranked according to their degree. | Possible hub role | MAX |
| S3 | Closeness | Closeness centrality focuses on how close a node is to all the other nodes in a network (Wasserman and Faust, 1994) and how long it will take to spread information from the node to all other nodes sequentially (Newman, 2005). | Distance | MIN |
| S4 | Betweenness | Betweenness represents the total amount of flow that a node carries when a unit of flow between each pair of nodes is divided up evenly over the shortest paths possible (Kleinberg and Easley, 2010). High betweenness nodes occupy critical roles in the network ("gatekeepers"). | Knowledge flow | MAX |
| S5 | Time sharing | The network configuration can provide information about possible working time needed to pass information to the node's neighbor. If the node is connected with a number of other nodes, its working time has to be divided and shared between all connected nodes. | Time sharing | MAX |
| S6 | Dissemination | Based on the information about our neighborhood (neighbors of our neighbors) we select the most linked nodes for future cooperation. In this strategy, we select the node with a lower degree, but one that is still connected to high-degree nodes. We focused on the potentially best-connected future source of knowledge. | Small world | MIN |

It is important to notice that strategies S3 and S4 are strongly dependent on the weights in the network beyond the topological effects. The relationship between nodes is weighted in proportion to the organization's structure at an organizational, social, and cognitive level. As a result, we have to use a weighted version of the algorithm to determine closeness and betweenness (Opsahl et al., 2010; Opsahl and Panzarasa, 2009).

Strategies S5 and S6 are based on the information about the nodes' neighborhood configuration that was reflected in the Co-Author Model (Tambayong, 2007). An important aspect of networks with multiple relations is the possibility of node cooperation time (S5). This function is understood as the ability of a node to make its resources available to other nodes. We can define the cooperation time based on the Co-Author Model. The Co-Author Model is a metaphor for the works of researchers who spend time writing papers. According to Jackson and Wolinsky (1996), a link represents the collaboration between two researchers, and the amount of time a researcher spends on any given project is inversely related to the number of projects that particular researcher is involved in. In this model, indirect connections will enter the utility function in a negative way, as they detract from one's coauthor time (Tambayong, 2007). The cooperation time strategy for node $i$ from network $N$ is formulated in the following way (Jackson and Wolinsky, 1996):

$$u_i(N) = \begin{cases} 0, if\, n_i = 0 \\ \sum_{ij \in N} \left[ \dfrac{1}{n_i} + \dfrac{1}{n_j} + \dfrac{1}{n_i n_j} \right] = 1 + \left(1 + \dfrac{1}{n_i}\right) \cdot \sum_{i \in N} \dfrac{1}{n_j} \,, if\, n_i > 0 \end{cases}$$

(6)

Function (6) represents the time, attention, and resources derived by $i$ from direct contact with $j$, when $i$ and $j$ are involved in $n_i$ and $n_j$ relations, respectively (Jackson and Wolinsky, 1996). The greatest value for the function is given to the node that works with many coworkers on an exclusive basis. On the other hand, the smallest value means that the node is connected with other nodes to a high degree. Such observations are the basis for strategy S6. Moreover, the weight between nodes does not affect the S5 and S6 strategies.

## 4.3  Simulation Results

The proposed model was verified during simulations in terms of knowledge diffusion and the development of competence within an organizational social network based on knowledge workers' collaborative learning. Simulations were performed on the Wats-Strogatz network with 0.1 rewiring probability and 484 nodes. Each network node was assigned an initial competence from the range (0, 10) using the previously defined competence vector $\overline{C}_i = [\, c_{1,i}, c_{2,i}, ...., c_{n,i}, ...., c_{N,i} \,]$ with ten elements and masked with binary values representing the availability to receive and transfer competence. The main goal of the simulations was to show the areas of application in competence management within an organization based on knowledge workers and their behaviors in the area of collaborative learning. Simulations were performed with the parameter $B = 0.006$, which represented the process of forgetting knowledge. During the simulations, the proposed strategies were verified for the selection of knowledge workers for specific roles such as experts, for the increased edge weights representing social relations, and for the knowledge collectors storing knowledge in the knowledge bank. In the first step, the role of experts within the network was modeled and the process of selection occurred based on six strategies (see Table 1). The results were compared with a reference simulation (R) based on the knowledge flow without identified roles. Using the proposed model, it was possible to simulate changes after increasing the competence of experts with knowledge randomly assigned from the range (10–50). Ten percent of the nodes were selected according to strategies (S1–S6) and the results were modeled in 500 steps. Fig. 2 presents the average competence from the simulation. Moreover, the reference simulation, without any changes, was added to the result shown in Fig. 2.

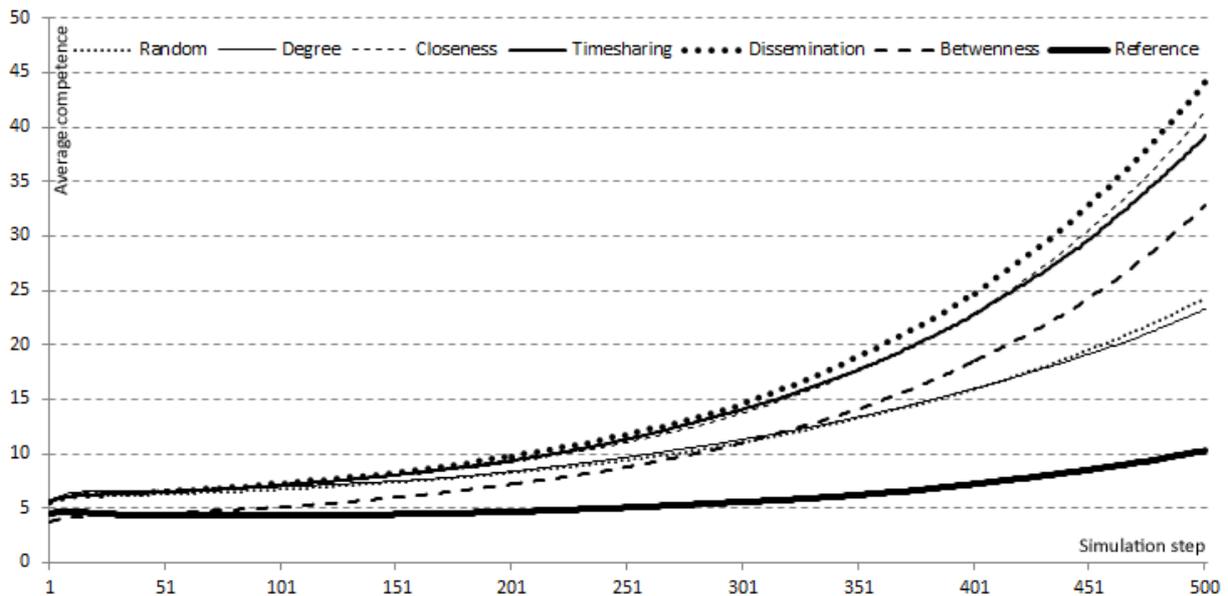

**Fig. 2.** Simulations based on increasing competence for selected nodes (by introducing experts to the network)

The initial starting competence resulted in an average value of five and was stabilized during the first 100 steps of the simulation. The best results were obtained for the dissemination strategy, with an average competence of 45 in the 500[th] step of the simulation. A strategy based on selecting knowledge workers with maximal closeness resulted in a 10% smaller result for average competence and was similar to a time sharing based strategy with an average competence of 40. The expert selection strategy, along with betweenness, resulted in a 20% smaller result with a 32.5 average value of competence, while the degree-based strategy was similar to a random strategy in its measurements.

The simulations represent a situation in an organization where there is a real need to increase the competence of a selected group of knowledge workers. One of the approaches is training, which generates additional costs. Another approach can be based on the knowledge facilitator, who is responsible for better communication and access to resources. This approach is based on increasing the weights representative of social relations for a selected set of nodes. The selection of nodes can be performed using different network measures (strategies S1–S6); the results are presented in Fig. 3.

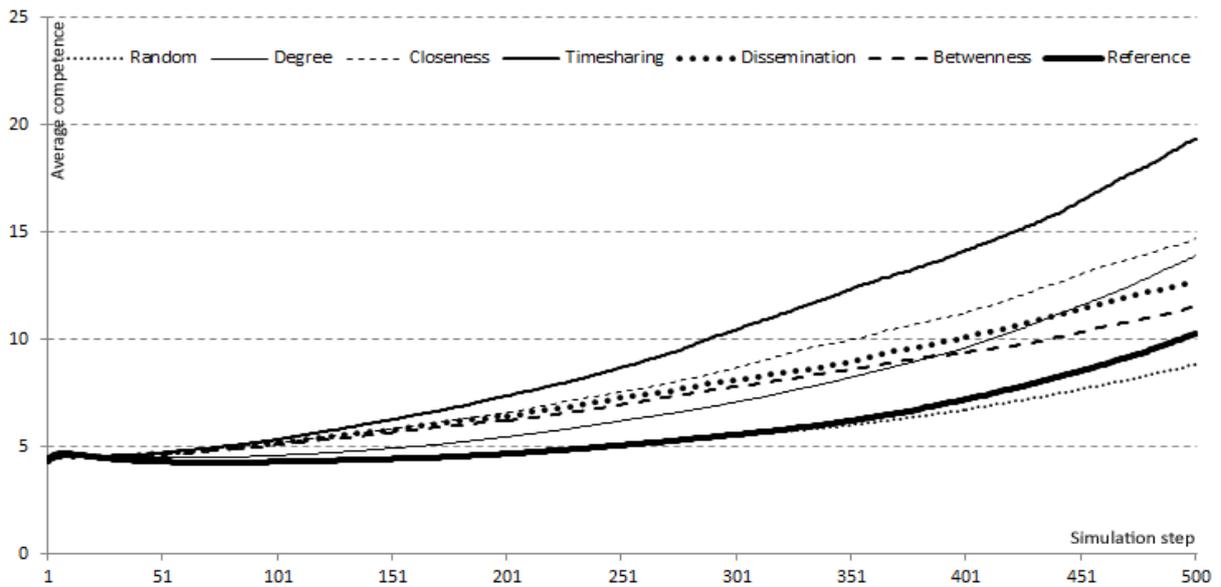

**Fig. 3.** Simulations based on increasing weights for selected nodes (by introducing a knowledge facilitator to the network)

Simulations were performed in 500 steps and the weights between the 50 selected nodes were increased by 20%. The highest average competence was obtained for the strategy based on sharing time and resulted in an average competence of 20. The strategy based on the selection of nodes with high closeness resulted in an average competence of 15. The degree-based strategy delivered results of 13.5 and outperformed the dissemination strategy by 10%. The betweenness-based strategy delivered an average competence of 11.5. The reference average competence based on simulations without changing the weights delivered similar results to random selection. Increasing the value of weights represents a situation within an organization where social relations can be improved and results in better knowledge flow.

Selecting simulated roles can improve the flow of knowledge within a network; for example, the role of a knowledge collector can improve competence management and the use of stored knowledge. Selecting workers responsible for knowledge collection can be done based on the strategies used for expert selection. This role can be assigned using the presented strategies; results are presented in Fig. 3 for the 50 collectors selected within the network.

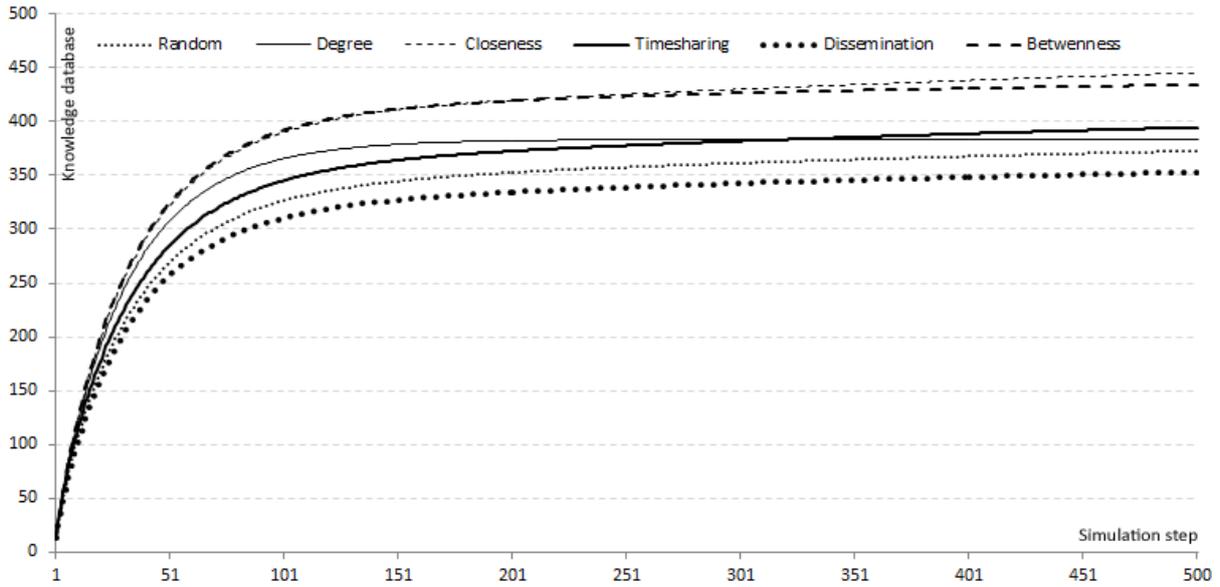

**Fig. 4.** Simulation based on the role of knowledge collector

During the simulations, the total gathered knowledge was computed and compared for different strategies. The best results were obtained for collectors based on the betweenness and closeness strategies. The aggregated value of competence for both strategies was 145 at the 500[th] step of the simulation. There was a 10% decrease in results with a value of only 400 for the strategy related to the measure of sharing time, closely followed by the degree strategy. The worst results were obtained for the dissemination strategy, with only a value of 350, which was 8% lower than the random selection strategy. For all strategies, the level of gathered knowledge stabilized after dynamic growth in the first 100 steps of the simulation.

The simulation shows that the best node for the expert role can be selected according to its neighborhood structure. This is because the main expert's role is to provide new knowledge in a network. In the first step the new knowledge is distributed to the expert's neighborhood. At this point it is important to accurately transfer as much knowledge as possible. In the next step nodes from the neighborhood redistribute the knowledge to their own neighborhoods based on their connections. Here, having a dense neighborhood structure is important. If we set aside the nodes' cognitive/social characteristics and knowledge potential, the most effective node for the expert role is the one with the most nested neighborhood. The best neighborhood structure for the expert role is a subject for future research. Moreover, interesting results may be gathered from clique analysis of sets of nodes from a node's neighborhood.

The second simulation approach (Fig. 3) focused on more effective knowledge distribution. The knowledge facilitator is selected to speed up the transfer of knowledge in certain parts of the network. The simulation shows that, similar to expert role selection, the knowledge facilitator selection process seeks nodes with the most efficient neighborhood. However, in this situation we focused on cliques that were explicitly separated from other parts of the network.

In the last simulation (Fig. 4) we looked for nodes with the best in/out transfer ratio in the network. The potential nodes for knowledge collection should play the role of transfer point in the overall network structure. Closeness and betweenness are best suited for the knowledge collector selection process because they take into consideration the overall network structure. Moreover, both of these metrics evaluate the value of relationships between nodes.

Generally speaking, interesting results regarding the presented tasks can be obtained if we analyze the nodes' cognitive/social characteristics and their knowledge level in addition to the network structure.

# 5  Community of Practice Acceleration

## 5.1  Knowledge Flow

The community of practice will be discovered through the analysis of the working area of each user (node selection vector $\bar{\alpha}^i$). More specifically, some parts of the selection vector are chosen and form the core for the community of practice $\bar{\alpha}_\zeta$, for $\bar{\alpha}_\zeta \subset \bar{\alpha}^i$. If the selection vectors are compatible, then we can assume that the related workers are working in the same area of interest and can be matched to the same community. Next, by analyzing knowledge flows we try to improve the effectiveness of each community of practice. Knowledge flow is the passing of knowledge between nodes according to certain rules and principles (Zhuge, 2006). In addition to knowledge flows, we analyzed the knowledge energy of each node in order to identify the importance of each node in the knowledge flow and the community. The node's knowledge energy is a numeric representation developed by Zhuge (2006) of each node's cognitive and creative ability. The knowledge energy is the power to drive knowledge flow, so it is also called "knowledge power" or "knowledge intensity" (Zhuge, 2004). Furthermore, in the proposed model, node $i$'s knowledge energy is estimated based on the node level of competence, as well as on its cognitive and social abilities according to the formula:

$$\hat{e}_i = (\overline{C}_i \cdot \overline{\alpha}_z)s_i o_i \qquad (7)$$

Formula (7) reflects the node's knowledge potential in a network for the community of practice $z$. From the dot product of competence and selection vectors, information about the importance of the community of practice and its levels is obtained. To formulate the full image of a node's knowledge potential, we should account for the node's ability to learn and teach as a base for knowledge transfer and assimilation in the knowledge flow.

In order to create effective knowledge flows, the following principles must be fulfilled (as defined by Zhuge et al. [2005]):

- Between any two nodes, knowledge only flows when their energies differ in at least one unit field.
- A knowledge flow network is efficient if every flow is from a node of higher energy to one of lower.
- Knowledge energy differences tend to diminish without reserve.
- If knowledge does not depreciate, then its energy will never decrease.

The presented principles provide some idea of how to manage knowledge flows in the community of practice. The most important statement is the one related to the order of nodes. In general, the knowledge flow should move from the node with the highest energy to a node with less energy, all the way down to the smallest node.

## 5.2 Community of Practice Acceleration Procedure

Due to network complexity, it is extremely difficult to develop methods for an optimal solution that can accelerate the community of practice's growth. The proposed procedure is a heuristics-based approach to the problem. The aim of the presented procedure for the community of practice's acceleration is to improve the knowledge flows between community members. In other words, the analysis of relationships between community members and node energy allows for decisions to be made about various ways to accelerate the community's knowledge flows. In the proposed approach, we improve the community knowledge flow transfers by creating new relationships between community members. We did not consider the problem of deleting relationships, as we cannot damage the existing structures in an organization.

The community of practice acceleration procedure starts with community detection. The detection process is based on the node selection vectors and looks for the community of practice core $\overline{\alpha}_z$. The selection vector for the community of practice core helps identify core community competences. We assume that nodes with similar selection vectors work in the same field of activity and use the same set of competences. Node classification is maintained by multi-label classification (Madjarov et al., 2012). As a result, the set of network nodes is divided into overlapping sets of nodes within communities $K = \bigcup\limits_{z=1}^{Z} k_z, k_z = \{v_t\}, t \in I$.

Another important concept is the efficiency of knowledge transfer between nodes. According to Zhuge (2005), the flow transports knowledge from nodes with higher knowledge energy to nodes with lower energy. The efficiency of knowledge transfer reflects the shortest path for the transfer of knowledge calculated based on the assumption that the efficiency $e_{v_{t1}, v_{t2}}$ of the tie between connected node $v_{t1}$ and any $v_{t2}$ is equal to $e_{v_{t1}, v_{t2}} = s_{t1} \cdot f(v_{t1}, v_{t2}) \cdot o_{t2}$. In other words, the relationship is influenced by associations with the starting nodes' social (teaching) abilities, the weight of the relationship itself, and the receiving nodes' cognitive (learning) abilities. The efficiency of knowledge transfer between any nodes from the community is calculated as:

$$\widetilde{e}_{v_x, v_y} = \left( \sum_{x=q}^{y=q-1} s_q \cdot f(v_q, v_{q+1}) \cdot o_{q+1} \right) \Big/ \widetilde{d} \Big/ \widetilde{d} \qquad (8)$$

where $\widetilde{d} = d - 1$, and $d$ is the number of nodes in the shortest path. $v_x$ is a starting node and $v_y$ is a final node in the path when $v_x, v_y \in k_z$. Moreover the shortest path is defined as an ordered set $Q = \{v_q\}$, where $v_q \in k_z$, and $(v_q, v_{q+1})$ are the subsequent pair of nodes in the shortest path set.

Our concept of community acceleration is related to a more efficient knowledge flow between nodes of a selected community. In order to accelerate the knowledge flow, the proposed procedure will suggest the location of a new tie and its value. The community of practice acceleration procedure is as follows:

1. Classify nodes in order to discover their communities $k_z, z = 1, 2, ..., Z$.

2. For every element (node) of the selected community $k_z$, calculate its knowledge energy $\forall v_t \in k_z : \hat{e}_t$ based on formula (7).

3. Order the nodes in the community set $(k_z, >)$ according to their energy $\hat{e}_t$. If $\exists v_{t1}, v_{t2} \in k_z : \hat{e}_{t1} = \hat{e}_{t2}$, then we make the next steps for $v_{t1}, v_{t2}$ separately.

4. Starting from the node with the highest energy calculated, the efficiency of knowledge transfer between ordered pairs of nodes $\widetilde{e}_{v_{t1}, v_{t2}}$ is determined based on formula (8).

5. A pair of nodes with the smallest value of efficiency of knowledge transfer is selected. A new direct tie between them is created. The strength of this tie can take different values. In our approach we heuristically assumed that the tie is equal to the average tie's strength in the network.

The presented procedure is applied to each community $k_z$ for a period of time in order to achieve the assumed efficiency of knowledge transfer between the nodes. On one hand, the procedure should be applied based on need due to the continually changing node energies. On the other hand, the procedures create ties that can be costly to maintain. In some cases, the new relationship creation idea is questionable due to worker differences in base knowledge or a different position in the company's structure.

## 5.3 Simulation Results

To illustrate the proposed approach in a detailed way, a Wats-Strogatz network with 0.1 rewiring probability and 25 nodes was generated. Each node was assigned an initial competence $c_i$ from the range (0, 10) and masks $m_j$ with binary values representing the availability to receive and transfer competences. The masks represent selection the vectors that are assigned to each node. The nodes were grouped into three clusters: $C_1$, $C_2$, and $C_3$ based on mask similarity, and a core set of identical competences with binary masks was identified for each cluster. For the first cluster $C_1$, the set of nodes [3, 5, 7, 11, 12, 13, 23] was assigned; cluster $C_2$ was assigned nodes [2, 4, 6, 8, 10, 14, 15, 16, 21, 22, 24], and cluster $C_3$ was assigned nodes [0, 1, 9, 17, 18, 19, 20]. Within the first cluster, competence $c_5$ with mask $m_5$ was identified as a core competence; for the second cluster, core competences are based on a set with masks $m_9$ and $m_{10}$. Within cluster three, a set of competences with masks $m_5$ and $m_7$ was identified as the core. The social network with the illustrated clusters based on competence vectors is shown in Fig. 5.

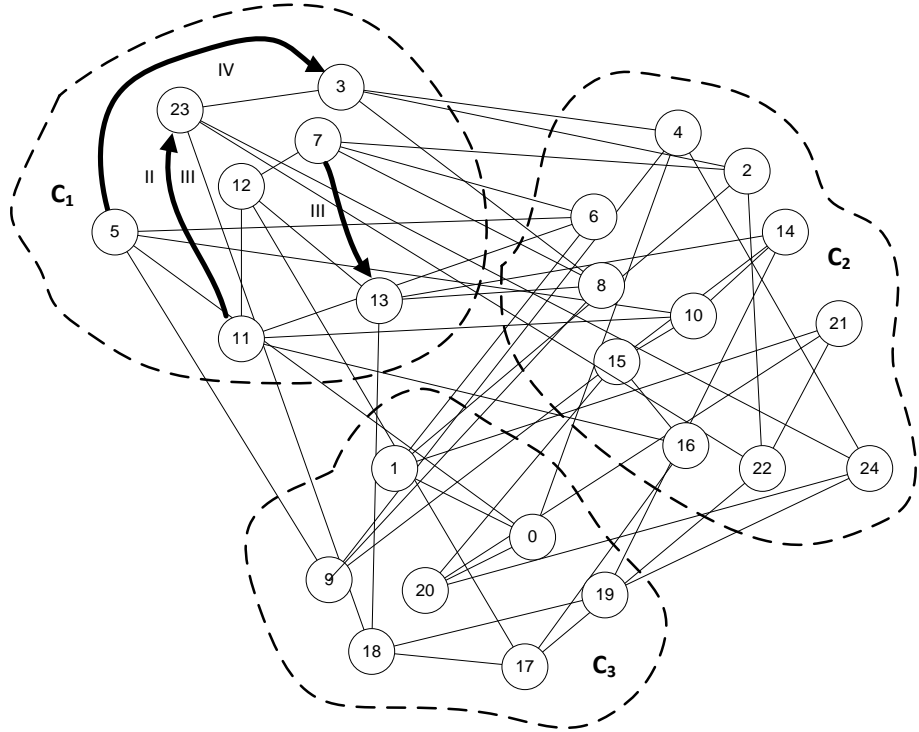

**Fig. 5.** Structure of social network and clusters based on competence vector

The problem within the organization can be related to the consideration that even knowledge workers are similar in terms of attributes; they can be unconnected or weakly connected to potential coworkers. In such situations, creating additional ties can improve the network's characteristics. Knowledge flow within networks was analyzed without any changes, and in the second step an additional random link was added within cluster $C_1$ between nodes 11 and 23 (II). In the next step, a second random connection between nodes 7 and 13 (III) was added. In the fourth step of the simulations, a connection was computed using the proposed approach and resulted in a connection between nodes 5 and 3 (IV).

Simulations were performed on four versions of the network in 500 steps to compare results within the network. The main goal of the simulations was to improve knowledge flow and monitor core competence, which was represented by mask $m_5$ for nodes within cluster $C_1$ with a set of nodes ($N_3$, $N_5$, $N_7$, $N_{11}$, $N_{12}$, $N_{13}$, $N_{23}$). The results of the simulations are presented in Fig. 6–Fig. 9.

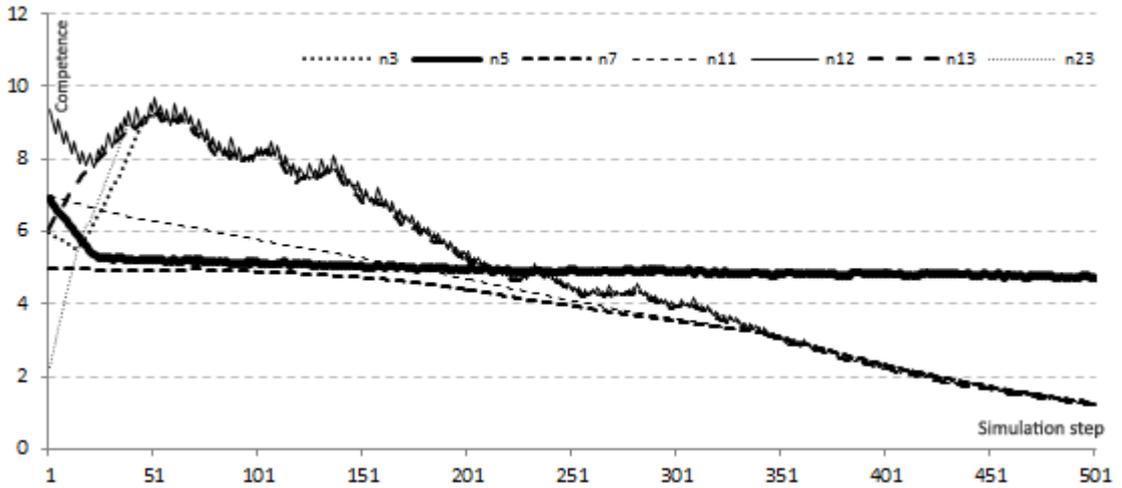

**Fig. 6.** Simulations based on a regular network

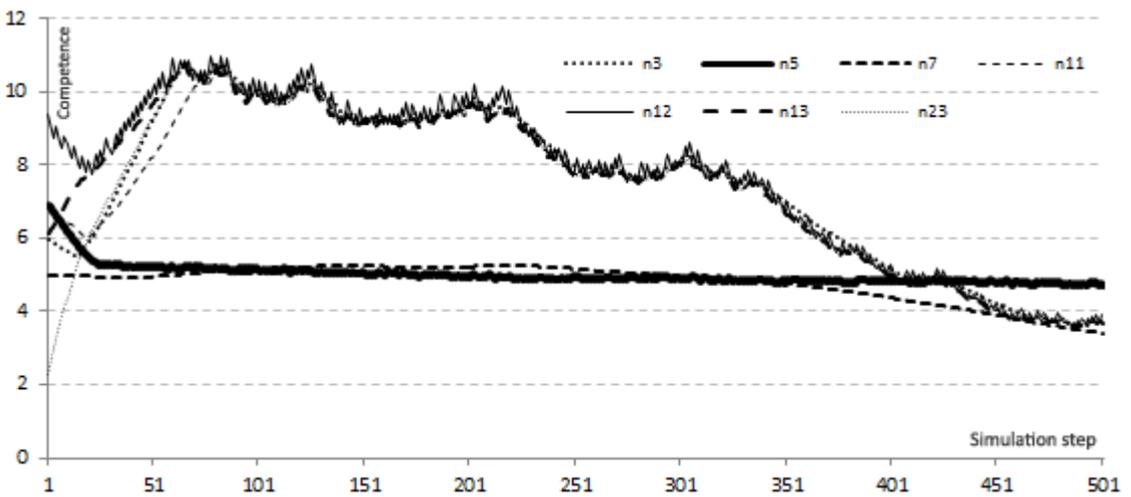

**Fig. 7.** Simulations based on a random link added from node $N_{11}$ to $N_{23}$

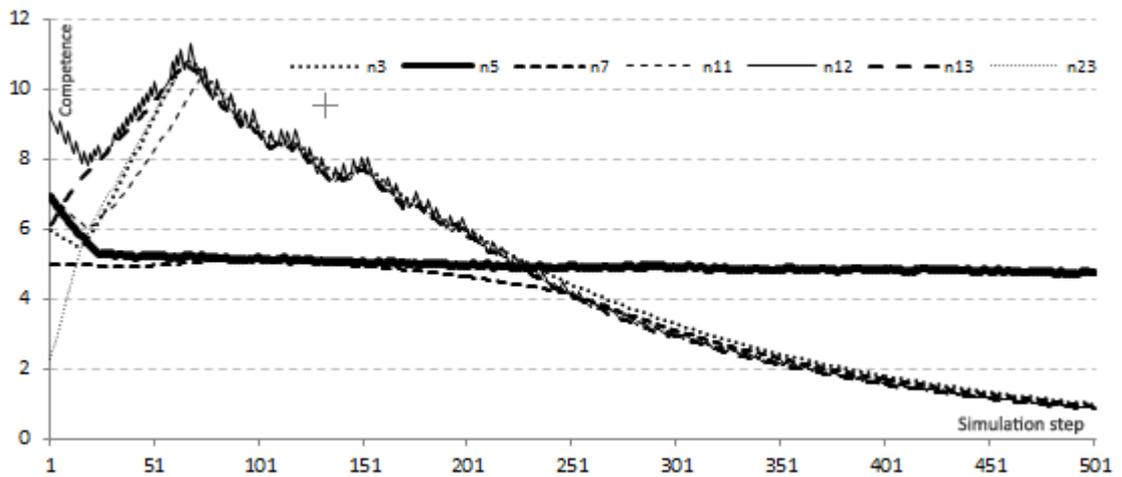

**Fig. 8.** Simulations based on two random links added from node $N_{11}$ to $N_{23}$
and from node $N_{12}$ to node $N_{13}$

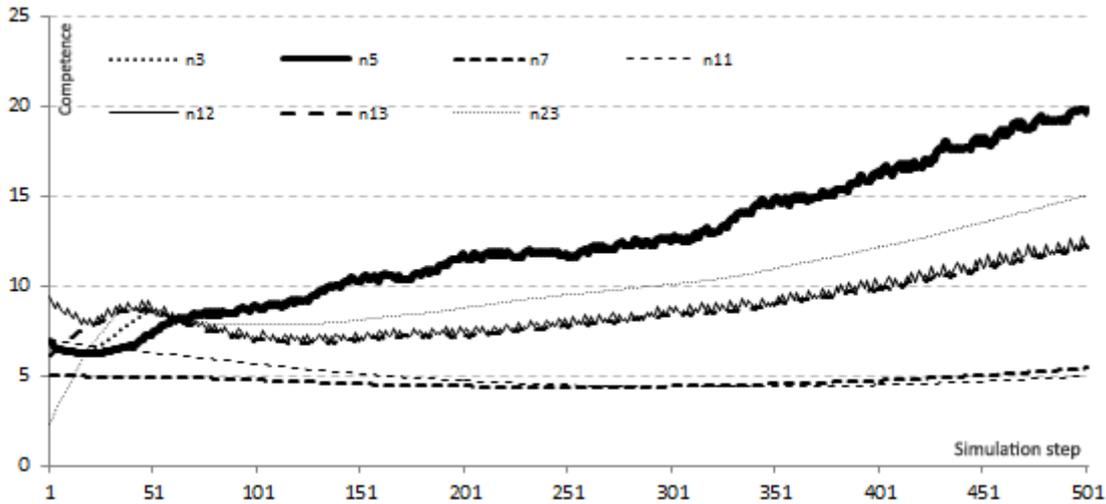

**Fig. 9.** Simulations based on link $N_3 \rightarrow N_5$ added using the proposed approach

Simulations performed on an unchanged network resulted in a maximal level of 9.5 for competence $c_5$ and then demonstrated a continuous drop, which is visible in Fig. 6. Adding single random links improved the results and a maximal level of 11 was obtained (Fig. 7). Similar results were achieved for a network with two random links added within the first cluster (Fig. 8). Even though maximal results were improved, there was still an observed drop in competence across the network. Use of the proposed method for the selection of new connections between nodes $N_3$ and $N_5$ is illustrated in Fig. 9. The proposed approach resulted in improvements within the cluster for most nodes and competence $c_5$ increased.

# 6   Conclusion

One of the important features of the proposed approaches is their ability to accurately predict organizational network development. In order to predict the knowledge flow movement we have to acquire information about worker competences and mutual relationships. The competence audit is a complex and costly operation. In normal conditions, an organization is able to maintain only a limited number of audits, usually once per year for each worker. For this reason, the ability to predict the future changes in an organizational network and worker competence level is very valuable. The presented approach, based on network behavior, allows the prediction of worker characteristics depending on worker roles, membership in communities of practice, and new relationships between the workers.

Moreover, the knowledge collector role facilitates the analysis of the company's repository development.

The knowledge workers' collaborative learning behavior model is based on knowledge flows and resource modeling. From the modeling side, the learning–teaching process is a complex activity where both sides have their own interests, which are reflected by their strategies. Generally, knowledge workers seek to transfer knowledge to other workers and follow organizational objectives in order to achieve some level of competence through them. The presented model helps analyze and change a given node's learning behaviors by changing competence levels or the tie structure in order to increase a company's (average) level of competences.

The results presented based on the simulations illustrate the effects of varied role allocations and strategy selections in the competence development process using the proposed model. The application of the model in management processes makes it possible to manage the development of competence within a given company. The presented simulations illustrate selected applications of the model. Depending on the structure of the network, different strategies for increasing competence within selected nodes can be assigned. Decisions can be made based on network measures, and prediction makes competence evaluation possible. Simulations make it possible to study the states of competence within the company without incurring costs related to continuous measurement or audit. Information about the structure of a network within the organization can be gathered from the analysis of email communications and can deliver useful assumptions and inputs for the model. The observations should be made to establish starting parameters for the model. The proposed approach makes it possible to track competences and observe the impact of connections on network performance in terms of knowledge diffusion and the role of network workers. Network performance with regard to knowledge flow can be improved by adding links within a community of practice.

Analyzing the structure of organizational social networks in terms of knowledge flow should be done in two stages, using the network structures and the attributes of the nodes. For future work, the proposed approach can be extended to the identification of communities within the graph and can seek to find relationships between clusters created with the vectors assigned to nodes; the results could then be verified using real-world datasets and more extensible simulations.

# 7 Acknowledgments

The work was partially supported by Fellowship co-Financed by European Union within European Social Fund, by European Union's Seventh Framework Programme for research, technological development and demonstration under grant agreement no 316097 [ENGINE] and by The National Science Centre, the decision no. DEC-2013/09/B/ST6/02317.